\documentclass[%
 reprint,
 amsmath,amssymb,
 aps,
longbibliography,
onecolumn,
notitlepage
]{revtex4-1}

\usepackage{graphicx}
\usepackage{dcolumn}
\usepackage{bm}
\usepackage{natbib}
\usepackage{color}

\begin{document}


\title{Mechanism of mean flow generation in rotating turbulence through inhomogeneous helicity}

\author{Kazuhiro Inagaki}
 \email{kinagaki@iis.u-tokyo.ac.jp}
\author{Nobumitsu Yokoi}%
\author{Fujihiro Hamba}%
\affiliation{%
 Institute of Industrial Science, University of Tokyo, Tokyo, Japan
}%

%


\date{\today}



 \makeatletter
  \renewcommand{\theequation}{\arabic{section}.\arabic{equation}}
  \@addtoreset{equation}{section}
 \makeatother

\allowdisplaybreaks[1]

\begin{abstract}
Recent numerical simulations showed that mean flow is generated in the inhomogeneous turbulence of an incompressible fluid that is accompanied by helicity and system rotation. In order to investigate the mechanism of the phenomenon, we perform a numerical simulation of inhomogeneous turbulence in a rotating system. In the simulation, an external force is applied to inject inhomogeneous turbulent helicity, and the rotation axis is perpendicular to the inhomogeneous direction. Mean velocity is set to zero in the initial condition of the simulation. The simulation results show that the mean flow directed to the rotation axis is generated and sustained only in the case with both the helical forcing and the system rotation. We investigate the physical origin of this flow-generation phenomenon by considering the budget of the Reynolds-stress transport equation. The results indicate that the pressure diffusion term significantly contributes to the Reynolds-stress equation and supports the generated mean flow. The results also reveal that a model expression for the pressure diffusion is expressed by the turbulent helicity gradient coupled with the angular velocity of the system rotation. This implies that inhomogeneous helicity plays a significant role in the generation of the large-scale velocity distribution in incompressible turbulent flows.
\end{abstract}

\pacs{Valid PACS appear here}
\maketitle


\section{\label{sec:level1}Introduction}


Helicity density (hereafter simply denoted as helicity) is defined as the inner product of velocity and vorticity and is known to play a crucial role termed as the $\alpha$ effect in the dynamo action in magnetohydrodynamics \cite{moffattbook}. 
In contrast, the role of helicity in neutral hydrodynamic turbulence is not clearly understood to-date. Studies on helicity can be divided into two categories, namely studies on the emergence of helicity and studies on the effects of helicity on the dynamics of turbulence. In the former, the rise of statistically significant helicity spectrum of homogeneous turbulence is never found in the absence of ad hoc initialization or forcing \cite{cj1989}. Conversely, helicity is known to emerge in rotating inhomogeneous turbulence such as a convection zone in a rotating sphere \cite{duarteetal2016,ssd2014} or a rotating inhomogeneous turbulence in which the rotation axis is parallel to the inhomogeneous direction \cite{gl1999,rd2014,kapylaetal2017}. Therefore, the key in the emergence of helicity corresponds to the inhomogeneity of rotating turbulence. In the latter, most studies focus on homogeneous turbulence and effects on energy cascade. A few studies revealed that helicity does not crucially influence hydrodynamic flows in the context of the energy cascade.
For example, with the aid of the eddy-damped quasi-normalized Markovian (EDQNM) approximation, Andr\'e and Lesieur \cite{al1977} showed that helicity does not affect the energy cascade once the inertial range is established. Rogers and Moin \cite{rm1987} numerically showed that the correlation between helicity and the dissipation rate of the turbulent kinetic energy is tenuous in homogeneous isotropic turbulence, homogeneous shear turbulence, and turbulent channel flow. Wallace \textit{et al.} \cite{wallaceetal1992} experimentally confirmed the correlation between helicity and the dissipation rate in a turbulent boundary-layer, a two-stream mixing-layer, and grid-flow turbulence. They concluded that there is a tenuous relationship between small dissipation rate and large helicity except in the shear flows.

In contrast, helicity is expected to be important in dynamics of mean flow generation in inhomogeneous turbulence. This point was first discussed by Yokoi and Yoshizawa \cite{yy1993} in terms of the closure scheme for the Reynolds-averaged Navier--Stokes (RANS) formulation. They suggested that the spatial gradient of helicity coupled with the vortical motion of fluid affects the Reynolds stress (velocity--velocity correlation) and diminishes the turbulent momentum transfer.
Recently, Yokoi and Brandenburg \cite{yb2016} numerically revealed that the mean flow is generated in a system with both inhomogeneous helicity and system rotation. This phenomenon can be explained with a model expression for the Reynolds stress obtained by Yokoi and Yoshizawa \cite{yy1993}.
Flow generation in the context of the large-scale flow instability was also discussed by Frisch \textit{et al.} \cite{aka}, and it is termed as the anisotropic kinetic alpha (AKA) effect. However, Yokoi and Brandenbrug \cite{yb2016} noted that the flow generation due to the inhomogeneous helicity is suitable for treating flows at high Reynolds number, such as astro/geophysical flows, while the AKA effect is valid only for flows at low Reynolds number. Thus the model proposed by Yokoi and Yoshizawa \cite{yy1993} involves general physics of fully developed inhomogeneous turbulence. However, the origin of the helicity effect on the Reynolds stress was not demonstrated based on the Reynolds-stress transport equation. In this sense, the manner in which helicity affects the Reynolds-stress evolution continues to be unclear. 

The Reynolds stress is typically modeled by the eddy-viscosity representation, which is one of the simplest models for the Reynolds stress. The eddy-viscosity model represents the momentum transfer enhanced by turbulence, and the effective viscosity is augmented by turbulent motions. Pope \cite{pope1975} obtained a nonlinear eddy-viscosity model for the Reynolds stress from the Reynolds-stress transport equation model of Launder \textit{et al.} \cite{lrr1975} by neglecting the diffusion effect. The fore-mentioned nonlinear eddy-viscosity models represented a considerable improvement relative to the conventional models. However, in some flows, the models continue to exhibit difficulties in terms of performance. A representative case in which the models do not work well is a swirling flow in a straight pipe \cite{steenbergen,kito1991}. In the flow, the mean axial velocity exhibits a dent profile in the center axis region of the pipe, and the dent profile is significantly more persistent in the downstream region than those predicted by the eddy-viscosity type models. 

Yokoi and Yoshizawa \cite{yy1993} applied the turbulence model with inhomogeneous helicity effect on the Reynolds stress to a swirling pipe flow and successfully reproduced the sustainment of the dent mean velocity. Another description of the effect of helicity on turbulence was constructed by Yoshizawa \textit{et al.} \cite{yoshizawaswirl2011}. They introduced a timescale of helical motion into the model and obtained good results in a swirling pipe flow. The results suggest the importance of helicity effect in describing the properties of swirling flows.
This helicity effect is also discussed in the context of the sub-grid scale (SGS) modeling in relation to the over-estimation of dissipation rate in the use of eddy-viscosity-type SGS stress models \cite{yy2017}.
However, the terms obtained in Yokoi and Yoshizawa \cite{yy1993} or Yoshizawa \textit{et al.} \cite{yoshizawaswirl2011} were not directly linked to the systematic modeling of Pope \cite{pope1975}. This is because the mechanism by which helicity affects the Reynolds stress is not fully known, and thus the helicity effect in the Reynolds-stress evolution is not explicitly considered. In order to reveal the helicity effect on the Reynolds stress, we investigate the physical origin of the effect at the level of the Reynolds-stress transport equation.

In this study, we perform a numerical simulation of a rotating inhomogeneous turbulence driven by a helical external forcing.
Although the mechanism of the helicity generation is important, this is not examined here. We impose the helicity by external forcing in the present study and focus on the effect of inhomogeneous helicity on the mean flow.
The flow configuration is similar to that used by Yokoi and Brandenburg \cite{yb2016}. It has two homogeneous directions and one inhomogeneous direction, and the rotation axis is perpendicular to the inhomogeneous direction. In the configuration the mean flow is expected to emerge in the rotation-axis direction.
This flow configuration is similar to the low-latitude region of rotating sphere in which turbulence is radially inhomogeneous and its rotation axis is mostly perpendicular to the inhomogeneous direction \cite{duarteetal2016,ssd2014}. We also conduct simulations in non-rotating and/or non-helical forcing cases to identify the condition for the mean-flow generation. The helicity effect is tested in relation to the Reynolds-stress transport equation, and the origin of the mean-flow generation is explored.

The rest of this study is organized as follows. Section~\ref{sec:level2} summarizes the relationship between the eddy-viscosity-type turbulence model and the transport equation for the Reynolds stress. The model for the Reynolds stress including the helicity effect derived by Yokoi and Yoshizawa \cite{yy1993} is also presented. Section~\ref{sec:level3} presents the numerical setup and the simulation results. We also discuss the origin of the helicity effect on the Reynolds stress. A comparison between our results and the model expression of the Reynolds stress with helicity is given in Sec.~\ref{sec:level4}. The conclusions are discussed in Sec.~\ref{sec:level5}.

\section{\label{sec:level2}Model representations of the Reynolds stress and helicity effect}

The Navier--Stokes equation and the continuity equation for an incompressible fluid in a rotating system are given respectively as follows:
\begin{align}
\frac{\partial u_i}{\partial t} & =
- \frac{\partial}{\partial x_j} u_i u_j - \frac{\partial p}{\partial x_i}
+ \nu \frac{\partial^2 u_i}{\partial x_j \partial x_j} + 2 \epsilon_{ij\ell} u_j \Omega^F_\ell + f_i, 
\label{eq:2.1} \\
\frac{\partial u_i}{\partial x_i} & = 0, 
\label{eq:2.2}
\end{align}
where $u_i$ denotes the $i$-th component of the velocity, $p$ the pressure divided by the fluid density with centrifugal force included, $\nu$ the kinematic viscosity, $\Omega^F_i$ the angular velocity of the system, $f_i$ the external force, and $\epsilon_{ij\ell}$ the alternating tensor. We decompose a physical quantity $q [= (u_i, p, f_i)]$ into mean and fluctuation parts as follows:
\begin{align}
q = Q + q', \ \ & Q = \left< q \right>,
\label{eq:2.3}
\end{align}
where $\left< \cdot \right>$ denotes an ensemble average. Substituting Eq.~(\ref{eq:2.3}) into Eqs.~(\ref{eq:2.1}) and (\ref{eq:2.2}), we obtain the mean field equations,
\begin{align}
\frac{\partial U_i}{\partial t} & =
- \frac{\partial}{\partial x_j} \left( U_i U_j + R_{ij} \right) - \frac{\partial P}{\partial x_i}
+ \nu \frac{\partial^2 U_i}{\partial x_j \partial x_j} + 2 \epsilon_{ij\ell} U_j \Omega^F_\ell + F_i, 
\label{eq:2.4} \\
\frac{\partial U_i}{\partial x_i} & = 0,
\label{eq:2.5}
\end{align}
where $R_{ij} (= \left< u_i' u_j' \right>)$ denotes the Reynolds stress. The only difference between Eqs.~(\ref{eq:2.1}) and (\ref{eq:2.4}) corresponds to the Reynolds stress. Thus, the Reynolds stress solely represents the effects of turbulent motion on the mean velocity. In order to close the system of Eqs.~(\ref{eq:2.4}) and (\ref{eq:2.5}), a model expression for the Reynolds stress is required.

\subsection{\label{sec:level2a}Relationship between model and transport equation for the Reynolds stress}

The simplest model for the Reynolds stress is the eddy-viscosity model that is expressed as follows:
\begin{align}
R_{ij} = \frac{2}{3} K \delta_{ij} - 2 \nu_T S_{ij},
\label{eq:2.6}
\end{align}
where $K (= \left<u_i' u_i' \right>/2)$ denotes the turbulent kinetic energy, $\nu_T$ the eddy viscosity, \linebreak $S_{ij} [= \left( \partial U_i / \partial x_j + \partial U_j / \partial x_i \right) / 2]$ the strain rate of the mean velocity, and $\delta_{ij}$ the Kronecker delta. The eddy-viscosity model is not just an empirical model but can be obtained from the fundamental equation, i.e., the Navier--Stokes equation. Specifically, the model expression for the Reynolds stress is closely related to the transport mechanism of the Reynolds stress. A systematic way to obtain the eddy-viscosity-type model from the Reynolds-stress transport equation may be summarized as follows \cite{pope1975,yoshibook}. The exact transport equation for the Reynolds stress is expressed as follows:
\begin{align}
\frac{\mathrm{D} R_{ij}}{\mathrm{D} t} & =
P_{ij} - \varepsilon_{ij} + \Phi_{ij} + \Pi_{ij} 
 + T_{ij} + D_{ij} + C_{ij} + F_{ij}, 
\label{eq:2.7}
\end{align}
where $\mathrm{D}/\mathrm{D} t = \partial / \partial t + U_\ell \partial / \partial x_\ell$ denotes the Lagrange derivative. Here, $P_{ij}$ denotes the production rate, $\varepsilon_{ij}$ the destruction rate, $\Phi_{ij}$ the pressure--strain correlation, $\Pi_{ij}$ the pressure diffusion, $T_{ij}$ the turbulent diffusion, and $D_{ij}$ the viscous diffusion, $C_{ij}$ the Coriolis effect, and $F_{ij}$ the external work. They are respectively defined as follows:
\begin{subequations}
\begin{align}
P_{ij} & = - R_{i\ell} \frac{\partial U_j}{\partial x_\ell} - R_{j\ell} \frac{\partial U_i}{\partial x_\ell}, 
\label{eq:2.8a} \\
\varepsilon_{ij} & = \left< 2 \nu s_{i\ell} \frac{\partial u_j'}{\partial x_\ell}
   + 2 \nu s_{j\ell} \frac{\partial u_i'}{\partial x_\ell} \right>, 
\label{eq:2.8b} \\
\Phi_{ij} & = 2 \left< p' s_{ij} \right>, 
\label{eq:2.8c} \\
\Pi_{ij} & = - \frac{\partial}{\partial x_j} \left< p' u_i' \right>
   - \frac{\partial}{\partial x_i} \left< p' u_j' \right>,
\label{eq:2.8d} \\
T_{ij} & = - \frac{\partial}{\partial x_\ell} \left< u_i' u_j' u_\ell' \right>, 
\label{eq:2.8e} \\
D_{ij} & = \frac{\partial}{\partial x_\ell} \left< 2 \nu s_{i\ell} u_j' + 2 \nu s_{j\ell} u_i' \right>, 
\label{eq:2.8f} \\
C_{ij} & = 2\left( \epsilon_{im \ell} R_{jm} + \epsilon_{jm \ell} R_{im} \right) \Omega_\ell^F ,
\label{eq:2.8g} \\
F_{ij} & = \left< u_i' f_j + u_j' f_i \right>,
\label{eq:2.8h}
\end{align}
\end{subequations}
where $s_{ij}  [= \left( \partial u_i / \partial x_j + \partial u_j / \partial x_i \right) / 2]$ denotes the strain rate of the velocity. Pope \cite{pope1975} obtained a general expression of the Reynolds stress based on the following two assumptions. First, to the right-hand side of Eq.~(\ref{eq:2.7}), the model by Launder \textit{et al.} \cite{lrr1975} (LRR model) is adopted; $\varepsilon_{ij}$ and $\Phi_{ij}$ are modeled as follows:
\begin{align}
\varepsilon_{ij} & = \frac{2}{3} \varepsilon \delta_{ij}, 
\label{eq:2.9} \\
\Phi_{ij} & = - C_{S1} \frac{\varepsilon}{K} B_{ij} + C_{R1} K S_{ij} \nonumber \\
& \hspace{1.2em} + C_{R2} \left[ B_{i\ell} S_{\ell j} + B_{j\ell} S_{\ell i} \right]_D 
 + C_{R3} \left( B_{i\ell} \Omega_{\ell j} + B_{j\ell} \Omega_{\ell i} \right) ,
\label{eq:2.10}
\end{align}
where $\varepsilon (= \varepsilon_{ii}/2)$ denotes the dissipation rate of the turbulent energy $K$, $B_{ij} [= R_{ij} -(2/3) K \delta_{ij}]$ the deviatoric part of the Reynolds stress, $\Omega_{ij} = \left( \partial U_j / \partial x_i - \partial U_i / \partial x_j \right)/2$, $\left[ A_{ij} \right]_D = A_{ij} - A_{\ell \ell} \delta_{ij} / 3$, and $C_{S1}$, $C_{R1}$, $C_{R2}$, and $C_{R3}$ denote the model constants.
The term with $C_{S1}$ describes the `return to isotropy' model while the terms with $C_{R1}$, $C_{R2}$, and $C_{R3}$ correspond to the `isotropization of production' model \cite{lrr1975}. Although there are more elaborate models for the pressure--strain correlation, such as Craft and Launder \cite{tcl}, we focus on simple models proportional to $B_{ij}$. Second, quasi-homogeneity of the flow field is assumed, and the diffusion terms are neglected as $\Pi_{ij} = T_{ij} = D_{ij} = 0$. In addition to the two assumptions, it is necessary to handle the time derivative term, $\mathrm{D} R_{ij} / \mathrm{D} t$. In the algebraic stress models, the weak-equilibrium assumption, $\mathrm{D} (R_{ij}/K)/\mathrm{D} t = 0$, is applied. Here this assumption is not used; we introduce an appropriate time derivative instead of the Lagrange derivative in order to satisfy the frame invariance of the turbulence equation in a rotating system \cite{hamba2006,ariki2015}. When the upper convected time derivative, $\mathfrak{D} A_{ij}/\mathfrak{D} t = \mathrm{D} A_{ij}/\mathrm{D} t - A_{i\ell} \partial U_j / \partial x_\ell - A_{j\ell} \partial U_i / \partial x_\ell$, is adopted, Eq.~(\ref{eq:2.7}) is expressed as follows:
\begin{align}
\frac{\mathfrak{D} B_{ij}}{\mathfrak{D} t} & =
 - C_{S1} \frac{\varepsilon}{K} B_{ij}
- \left( \frac{4}{3} - C_{R1} \right) K S_{ij} \nonumber \\
& \hspace{1.2em}
- \left( 1 - C_{R2} \right) \left[ B_{i\ell} S_{\ell j} + B_{j\ell} S_{\ell i} \right]_D 
- \left( 1 - C_{R3} \right) \left( B_{i\ell} \Omega_{\ell j}^* + B_{j\ell} \Omega_{\ell i}^* \right),
\label{eq:2.11}
\end{align}
where $\Omega_{ij}^* (= \Omega_{ij} + \epsilon_{ij\ell} \Omega_\ell^F)$ denotes the mean absolute vorticity tensor. Here it is assumed that the external work does not affect the Reynolds stress directly. The model for $\Phi_{ij}$ is extended to a rotating system. Thus, we replace $\Omega_{ij}$ in Eq.~(\ref{eq:2.10}) by $\Omega_{ij}^*$. This frame invariant formulation is performed to ensure the consistency of the equations in a rotating frame. The effect of rotation may affect the transport equation for $\varepsilon$ \cite{bfr1985}, and this type of a modification is needed to predict turbulent flows under the solid body rotation with the Reynolds-stress models. However, this point is beyond the scope of the present study that focuses on the effects on the mean flow. The first term on the right-hand side of Eq.~(\ref{eq:2.11}) represents the destruction of $B_{ij}$ or the relaxation to an isotropic state. The second term denotes the production of $B_{ij}$ by the isotropic part of turbulence, while the third and fourth terms  denote the production by the anisotropic part of turbulence. Equation~(\ref{eq:2.11}) is re-expressed as follows:
\begin{align}
B_{ij} & =
- 2 \frac{4-3C_{R1}}{6C_{S1}} \frac{K^2}{\varepsilon} S_{ij} \nonumber \\
& \hspace{1.2em}
- \frac{1-C_{R2}}{C_{S1}} \frac{K}{\varepsilon} \left[ B_{i\ell} S_{\ell j} + B_{j\ell} S_{\ell i} \right]_D 
- \frac{1-C_{R3}}{C_{S1}} \frac{K}{\varepsilon} \left( B_{i\ell} \Omega_{\ell j}^* + B_{j\ell} \Omega_{\ell i}^* \right) 
\nonumber \\
& \hspace{1.2em}
- \frac{1}{C_{S1}} \frac{K}{\varepsilon} \frac{\mathfrak{D} B_{ij}}{\mathfrak{D} t}.
\label{eq:2.12}
\end{align}
Substituting this expression iteratively into $B_{ij}$ on the right-hand side, we obtain the following:
\begin{align}
B_{ij} & =
- 2 C_\nu \frac{K^2}{\varepsilon} S_{ij} \nonumber \\
& \hspace{1.2em}
+ C_{q1} \frac{K^3}{\varepsilon^2} \left[ S_{i\ell} S_{\ell j} + S_{j\ell} S_{\ell i} \right]_D 
+ C_{q2} \frac{K^3}{\varepsilon^2} \left( S_{i\ell} \Omega_{\ell j}^* + S_{j\ell} \Omega_{\ell i}^* \right)
\nonumber \\
& \hspace{1.2em}
+ C_d \frac{K}{\varepsilon} \frac{\mathfrak{D}}{\mathfrak{D} t} \left( \frac{K^2}{\varepsilon} S_{ij} \right) 
+ \text{(higher order terms)} , 
\label{eq:2.13}
\end{align}
where $C_\nu = (4-3C_{R1})/(6C_{S1})$, $C_{q1} = 2 C_\nu (1-C_{R2}) /C_{S1}$, $C_{q2} = 2 C_\nu (1-C_{R3})/C_{S1}$, and $C_d = 2C_\nu/C_{S1}$. In contrast to the formulation obtained by Pope \cite{pope1975}, the time derivative term is retained in the right-hand side of Eq.~(\ref{eq:2.13}) as shown in Yoshizawa \cite{yoshibook}. This corresponds to a more general formulation when compared with that obtained by Pope \cite{pope1975} since the time derivative term does not always disappear. The first term of Eq.~(\ref{eq:2.13}) represents the eddy-viscosity term which corresponds to the second term on the right-hand side of Eq.~(\ref{eq:2.6}), and this term is derived from the isotropic part of the production term. This reflects the point that the eddy-viscosity model constitutes a good approximation when the turbulence is nearly isotropic, quasi-homogeneous, and steady.

\subsection{\label{sec:level2b}The Reynolds-stress expression accompanied with the helicity effect}

The eddy-viscosity-type models provide good results for simple flows such as free shear layer flows and channel flows. However, they perform poorly for more complex flows. An example in which the usual eddy-viscosity models do not work well is a swirling flow in a straight pipe \cite{steenbergen,kito1991}. In the swirling-flow experiments, it is observed that the mean axial velocity shows a dent in the center axis region, and this inhomogeneous velocity profile is very persistent to the well downstream region. However, this type of a dent profile that is imposed at the pipe inlet cannot be sustained and decays rapidly in the usual eddy-viscosity model simulation \cite{ky1987,steenbergen}. This is because the eddy viscosity is so strong that it smears out any large-scale velocity gradient. Jakirli\'c \textit{et al.} \cite{jht2000} pointed out that even with an elaborate explicit Reynolds-stress model such as Craft \textit{et al.} \cite{cls1996} or Shih \textit{et al.} \cite{shihetal1997} as well as the standard eddy-viscosity model, it is difficult to accurately reproduce the fore-mentioned rotational flows without performing a few modifications in the model constants. With the aid of the two-scale direct-interaction approximation (TSDIA) \cite{tsdia} that is an analytical statistical theory of inhomogeneous turbulence, Yokoi and Yoshizawa \cite{yy1993} suggested that eddy viscosity may be suppressed by symmetry breaking swirling motion. They analytically constructed a new turbulence model in which the helicity effect is incorporated. In the formulation, homogeneous isotropic non-mirror-symmetric turbulence is assumed as the basic field, and the effects of inhomogeneity, anisotropy, and system rotation are incorporated in a perturbational manner based on the Navier--Stokes equation. Brief descriptions of the formulation are given in Appendix~\ref{sec:a}. According to the formulation, the deviatoric or traceless part of the Reynolds stress is expressed as follows:
\begin{align}
B_{ij} = - 2 \nu_T S_{ij}
+ \eta \left[ \frac{\partial H}{\partial x_j} \Omega^*_i + \frac{\partial H}{\partial x_i} \Omega^*_j \right]_D,
\label{eq:2.14}
\end{align}
where $\eta$ denotes the transport coefficient, $H (= \left< u_i' \omega_i' \right>)$ the turbulent helicity, and $\Omega^*_i (= \epsilon_{ij\ell} \partial U_\ell / \partial x_j + 2 \Omega^F_i)$ the mean absolute vorticity. In this study, we refer to the model of Eq.~(\ref{eq:2.14}) as the helicity model. This model allowed the successful reproduction of the sustainment of the dent mean axial velocity in a swirling flow. The helicity model is similar to the AKA model \cite{aka} in the sense that the AKA describes the effect of lack of parity invariance on the mean flow. The helicity model is developed for high-Reynolds number flows since the TSDIA corresponds to perturbational expansion from fully developed homogeneous turbulence, while the AKA is valid for low-Reynolds number flows \cite{yb2016}. Hence, it is expected the helicity model can be applied to realistic high-Reynolds number turbulent flows.

It is interesting to note that as pointed out in \cite{yy1993} and \cite{yb2016}, the present model accounts for the mean flow generation from the no-mean-velocity initial condition. Even if system does not have the mean velocity gradient, Eq.~(\ref{eq:2.14}) may include a non-zero value when both the helicity gradient and the system rotation exist. In such cases, the deviatoric part of the Reynolds stress is expressed as follows:
\begin{align}
B_{ij} = 2 \eta \left[ \frac{\partial H}{\partial x_j} \Omega^F_i + \frac{\partial H}{\partial x_i} \Omega^F_j \right]_D 
\neq 0.
\label{eq:2.15}
\end{align}
This suggests that the mean flow is generated by this helicity effect when the inhomogeneous helicity is coupled with the rotation since the mean velocity equation is expressed as
\begin{align}
\frac{\partial U_i}{\partial t} = - \frac{\partial}{\partial x_j} 
\left[ \eta \left( \frac{\partial H}{\partial x_j} 2 \Omega^F_i 
+ \frac{\partial H}{\partial x_i} 2 \Omega^F_j - \frac{\partial H}{\partial x_\ell} 2 \Omega^F_\ell \frac{2}{3} \delta_{ij} \right) \right] - \frac{\partial P}{\partial x_i}
\neq 0.
\label{eq:2.16}
\end{align}
Yokoi and Brandenburg \cite{yb2016} performed direct numerical simulations (DNSs) of a rotating inhomogeneous turbulence with an imposed turbulent helicity. They commenced with a no-mean-velocity configuration and observed a mean-flow generation in a rotating turbulence. Additionally, they confirmed that in the early stage of the simulation in which the mean-velocity gradient is not significantly developed, the Reynolds stress is well correlated with the middle part of Eq.~(\ref{eq:2.15}). It is not possible to predict this type of a flow generation phenomenon by using a conventional model of the Reynolds stress as given by Eq.~(\ref{eq:2.13}) since each term contains the mean shear rate.

The results indicate that inhomogeneous helicity coupled with the vortical motion of fluid affects the Reynolds stress and reduces turbulent momentum transport represented by the eddy viscosity. The following points should be noted. The model representation of Eq.~(\ref{eq:2.14}) was analytically obtained from the Navier--Stokes equation with the aid of TSDIA. However, the second term on the right-hand side of Eq.~(\ref{eq:2.14}) is not obtained in a direct manner from the systematic construction of the model shown in Sec.~\ref{sec:level2a}. This is because the turbulent helicity is not explicitly included in the Reynolds-stress transport equation given in Eq.~(\ref{eq:2.11}) on which the model constitution is based. Yokoi and Brandenburg \cite{yb2016} compared the profile of the Reynolds stress with that of Eq.~(\ref{eq:2.14}) to determine a very good correlation between them. However, the origin of the helicity effect on the Reynolds-stress equation was not shown. As shown in Sec.~\ref{sec:level2a}, the model expression of the Reynolds stress is related to  its transport mechanism. The effect of helicity corresponding to the second term on the right-hand side of Eq.~(\ref{eq:2.14}) should exist on the right-hand side of Eq.~(\ref{eq:2.7}) as well as the production term corresponding to the eddy-viscosity term. Hence, the physical origin of the second term of Eq.~(\ref{eq:2.14}) is not clarified in the sense  of the Reynolds-stress evolution.

\section{\label{sec:level3}Numerical simulations}

In order to investigate the mechanism of the mean-flow generation and its relationship to the turbulent helicity, we perform a series of numerical simulations of a rotating inhomogeneous turbulence driven by a helical external force. We examine the transport equation for the Reynolds stress to explore the manner in which the turbulent helicity affects the Reynolds-stress transport.

\subsection{\label{sec:level3a}Governing equations and numerical setup}

In order to simulate a high-Reynolds-number turbulent flow, the large eddy simulation (LES) is adopted instead of the DNS. The governing equations of the LES in a rotating system are expressed as follows:
\begin{align}
\frac{\partial \overline{u}_i}{\partial t} & =
 - \frac{\partial}{\partial x_j} \overline{u}_i \overline{u}_j
  - \frac{\partial \overline{p}}{\partial x_i}
 + \frac{\partial}{\partial x_j} 2 \nu_{sgs} \overline{s}_{ij} 
+ 2 \epsilon_{ij\ell} \overline{u}_j \Omega_\ell^F
  + \overline{f}_i , 
\label{eq:3.1} \\ 
\frac{\partial \overline{u}_i}{\partial x_i} & = 0,
\label{eq:3.2}
\end{align}
where the kinematic viscosity is neglected, and $\overline{q}$ denotes the grid-scale (resolved) component of $q$. It should be noted that $\overline{q}$ is different from the ensemble average, $\left< q \right>$, which is already introduced in Eq.~(\ref{eq:2.3}). With respect to the model of the subgrid-scale (SGS) viscosity, $\nu_{sgs}$, the Smagorinsky model \cite{smagorinsky1963},
\begin{align}
\nu_{sgs} = \left( C_S \Delta \right)^2 \sqrt{ 2 \overline{s}_{ij} \overline{s}_{ij}}, 
\label{eq:3.3}
\end{align}
is applied with the Smagorinsky constant $C_S = 0.19$, which is the optimized value for homogeneous isotropic turbulence \cite{yoshibook}, and $\Delta = \left( \Delta x \Delta y \Delta z \right)^{1/3}$ where $\Delta x_i$ denotes the grid size of the $i$-th direction.

\begin{figure}[b]
\centering
\includegraphics[scale=0.45]{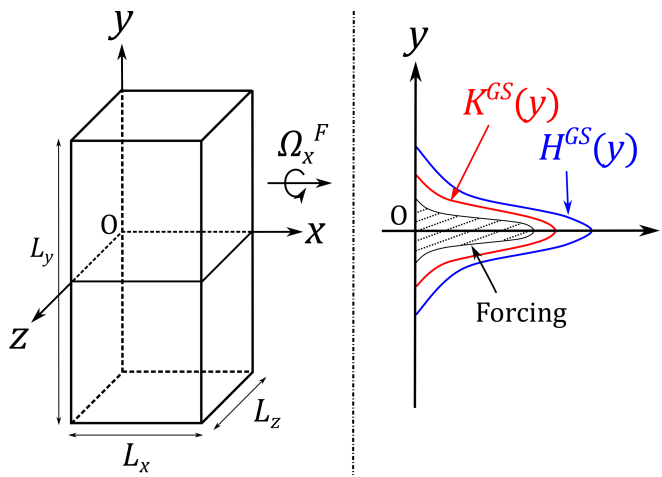}
\caption{Computational domain and schematic profiles of turbulent energy and helicity. $K^{GS} (= \left< \overline{u}_i' \overline{u}_i' \right>/2)$ and $H^{GS} (= \left< \overline{u}_i' \overline{\omega}_i' \right>)$ denote the turbulent energy and helicity of the grid scale motions, respectively. An external forcing is applied only around $y=0$ plane.}
\label{fig:1}
\end{figure}

In the simulation, the computational domain is a rectangular parallelepiped region as shown in Fig.~\ref{fig:1}. An external force applied around the center plane at $y=0$ injects turbulent energy and helicity. In the calculation, the rotation axis is set perpendicular to the inhomogeneous direction of the turbulence to assess the helicity model Eq.~(\ref{eq:2.14}). This set up is similar to that used by Yokoi and Brandenburg \cite{yb2016}. The configuration corresponds to the low-latitude region of a rotating  spherical convection in which the inhomogeneous direction of helicity is mainly perpendicular to the rotation axis in a low-latitude region \cite{duarteetal2016,ssd2014}. The objective involves elucidating the effect of inhomogeneous helicity on the mean flow in rotating turbulence and not clarifying the mechanism of helicity generation, and thus helicity is injected by an external forcing to achieve simplicity in contrast to the simulation in which helicity emerges spontaneously \cite{gl1999,rd2014,kapylaetal2017}. The external force is defined by the vector potential $\overline{\psi}_i$ as follows:
\begin{align}
\overline{f}_i = C \epsilon_{ij\ell} \frac{\partial}{\partial x_j} \left[ g (y) \overline{\psi}_\ell \right],
\label{eq:3.4}
\end{align}
where $g(y)$ denotes a weighting function introduced to confine the external force around the $y=0$ plane. The coefficient $C$ is determined to satisfy $\left< \overline{u}_i' \overline{u}_i' \right>_S (y=0)/2=1$ at each time step, where $\left< \cdot \right>_S$ denotes the $x$--$z$ plane average and $\overline{q}'$ denotes the fluctuation of $\overline{q}$ around $\left< \overline{q} \right>_S$; 
\begin{align}
\overline{q} = \left< \overline{q} \right>_S + \overline{q}'. 
\label{eq:3.5}
\end{align}
The force is solenoidal, $\partial \overline{f}_i / \partial x_i = 0$. With respect to the weighting function, $g(y) = \mathrm{exp} \left[ -y^2/\sigma^2 \right]$ with $\sigma = L_y/32 = 0.393$ is applied, and this is a value comparable to the forcing scale $\pi/k_f$ where $k_f$ is given in the following [Eq.~(\ref{eq:3.6a})]. The vector potential $\overline{\psi}_i$ obeys a stochastic process like the Ornstein--Uhlenbeck process \cite{ouforcing}, and is determined from the power and helicity spectra of $\overline{f}_i$, $E^{ex}(k)$ and $E_H^{ex}(k)$ given as follows:
\begin{subequations}
\begin{align}
E^{ex} (k) & \propto
\begin{cases}
k^{-5/3} & k=k_f, 10 \le k_f \le 14 \\
0 & \text{otherwise} ,
\end{cases} 
\label{eq:3.6a} \\
E_H^{ex} (k) & = 2 \alpha k E^{ex} (k) ,
\label{eq:3.6b}
\end{align}
\end{subequations}
where $\alpha$ denotes the parameter that determines the intensity of helicity of the external force. The spectrum $E^{ex}(k)$ is selected corresponding to the typical inertial-range form of turbulence, and $E^{ex}_H (k)$ corresponds to the statistical property of inertial wave when $\alpha = \pm 1$ \cite{moffatt1970}. The range of $\alpha$ should be $-1 \le \alpha \le 1$ since the helicity spectrum must satisfy $|E_H^{ex}(k)| \le 2k E^{ex}(k)$ \cite{al1977}; $\alpha = 0$ corresponds to the non-helical case and $\alpha = 1$ ($-1$) is the most positively (negatively) helical case. Details of forcing are given in Appendix~\ref{sec:b}.

The size of the computational domain is $L_x \times L_y \times L_z = 2\pi \times 4\pi \times 2\pi$ and the number of the grid point is $N_x \times N_y \times N_z = 128 \times 256 \times 128$. The periodic boundary conditions are used in all directions, we use the second-order finite-difference scheme in space, and the Adams--Bashforth method is used for time integral. A triply periodic box is used, and thus the pseudo-spectral scheme may be more appropriate for DNS with the linear viscosity term. However, with respect to the LES, a complex nonlinear form of the SGS viscosity decreases the numerical accuracy of the pseudo-spectral scheme. Moreover, we are going to apply the code to homogeneous turbulence with a non-uniform grid. Thus, we adopt finite-difference scheme. The pressure is directly solved in the wave number space by using FFT. Parameters of the simulation are shown in Table~\ref{tb:1}; namely Run 1 is non-helical and non-rotating case, Run 2 is helical but non-rotating, Run 3 is rotating but non-helical, and Runs 4, 5, and 6 are helical and rotating. We observe the effect of helical forcing by comparing Runs 1 and 2 for non-rotating case, and Runs 3, 4, and 6 for the rotating case. We also observe the effect of the system rotation by comparing Runs 2, 5, and 6. In all the runs, the external force is applied in the wavenumber band $10 \le k \le 14$. With respect to the helical cases, $\alpha = 0.5$ for Runs 2 and 4 and $\alpha = 0.2$ for Run 5, which are not fully helical ones, are adopted since the relative helicity [$u_i \omega_i/(|u_i||\omega_i|)$] in realistic turbulence is modulated from the maximally helical case of the inertial wave, $\alpha = \pm 1$, due to buoyancy and nonlinear interaction of turbulence \cite{ranjan2017}. $L^{GS}_0$ denotes the characteristic length scale of the turbulence and $\mathrm{Ro}^{GS}_0$ denotes the Rossby number respectively defined by
\begin{align}
L^{GS}_0 = \frac{(K^{GS}_0)^{3/2}}{\varepsilon^{SGS}_0} \ , \ \ 
\mathrm{Ro}^{GS}_0 = \frac{{K^{GS}_0}^{1/2}}{L^{GS}_0 2 \Omega^F} ,
\label{eq:3.7}
\end{align}
where $K^{GS} = \left< \overline{u}_i' \overline{u}_i' \right>/2$, $K^{GS}_0 = K^{GS} (y=0)$, $\varepsilon^{SGS} = 2\left< \nu_{sgs} \overline{s}_{ij} \overline{s}_{ij}' \right>$, $\varepsilon^{SGS}_0 = \varepsilon^{SGS} (y=0)$, and $\left< \cdot \right>$ denotes the average over the homogeneous plane and over time. The time average is taken over $20 \le t \le 30$ as mentioned below. In our calculation, the length scale of inhomogeneity of turbulence is estimated as $\ell^\nabla = 0.4$ for all runs, in which $\ell^\nabla$ is defined such that $K^{GS} (y= \ell^\nabla) = \mathrm{e}^{-1} K^{GS}_0$. The validity of turbulence models requires the length scale of inhomogeneity of turbulence is much longer than the scale of energy containing eddy $L_0^{GS}$ \cite{corrsin1974}. These two scales are comparable in the simulation. However, the fore-mentioned lack of scale separation is often observed in actual turbulence such as in an atmospheric boundary layer \cite{stull1993}. It should be emphasized that the mean velocity is set to zero in the initial condition, and the plane average of the external force is also zero, $\left< \overline{f}_i \right>_S = 0$, such that the external force does not directly excite the mean velocity.

\begin{table}[ht]
\centering
\caption{Calculation parameters.}
\begin{ruledtabular}
\begin{tabular}{ccccc}
Run & $\alpha$ & $\Omega_x^F$ & $L^{GS}_0$ & $\mathrm{Ro}^{GS}_0$ \\ \hline
1 & $0$ & $0$ & $0.506$ & $\infty$ \\
2 & $0.5$ & $0$ & $0.547$ & $\infty$ \\
3 & $0$ & $5$ & $0.542$ & $0.185$ \\
4 & $0.2$ & $5$ & $0.550$ & $0.182$ \\
5 & $0.5$ & $2$ & $0.544$ & $0.459$ \\
6 & $0.5$ & $5$ & $0.602$ & $0.166$ \\
\end{tabular}
\end{ruledtabular}
\label{tb:1}
\end{table}

\subsection{\label{sec:level3b}Numerical results}

\subsubsection{\label{sec:level3b1}Mean-flow generation}

\begin{figure}[htp]
 \begin{tabular}{c}
  \begin{minipage}{0.49\hsize}
   \centering
   \includegraphics[scale=0.74]{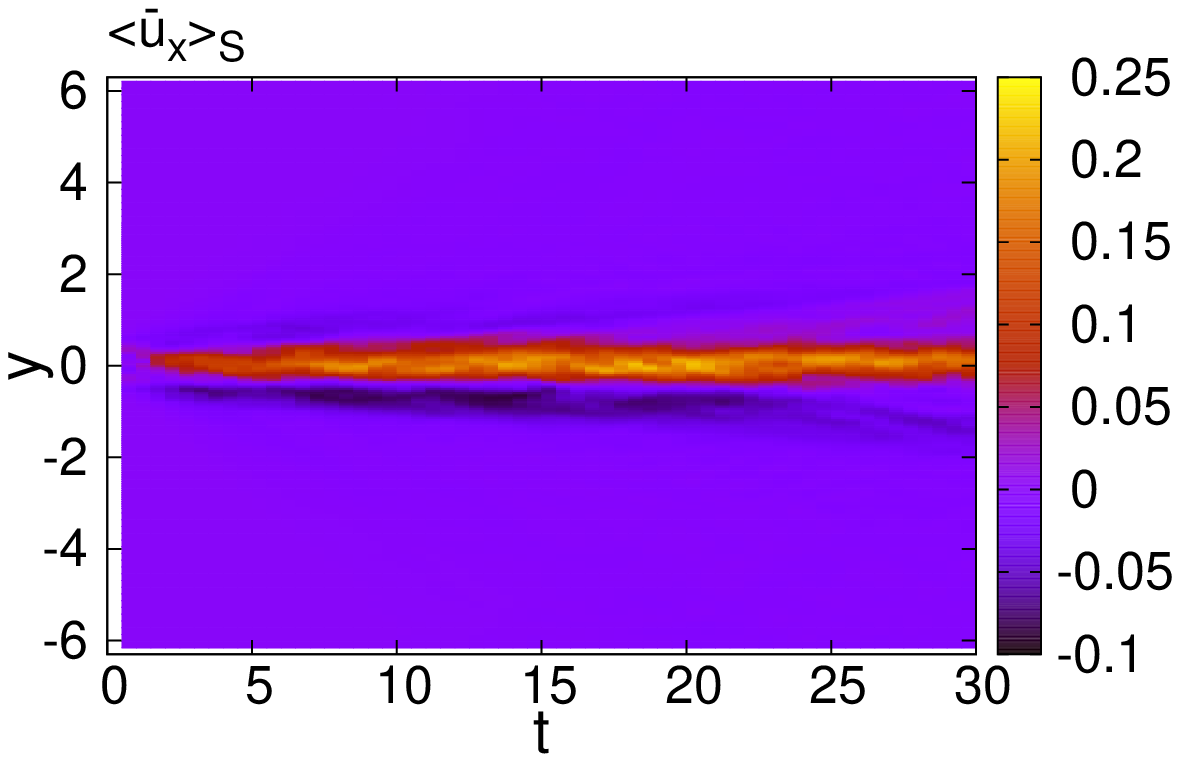}
   \caption{Time evolution of the axial mean velocity for Run 6. The horizontal axis denotes the time, and the vertical axis denotes the inhomogeneous direction, $y$, and the color contour denotes the value of $\left< \overline{u}_x \right>_S$.}
  \label{fig:2}
  \end{minipage}
  \begin{minipage}{0.04\hsize}
  \end{minipage}
  \begin{minipage}{0.49\hsize}
   \centering
   \includegraphics[scale=0.67]{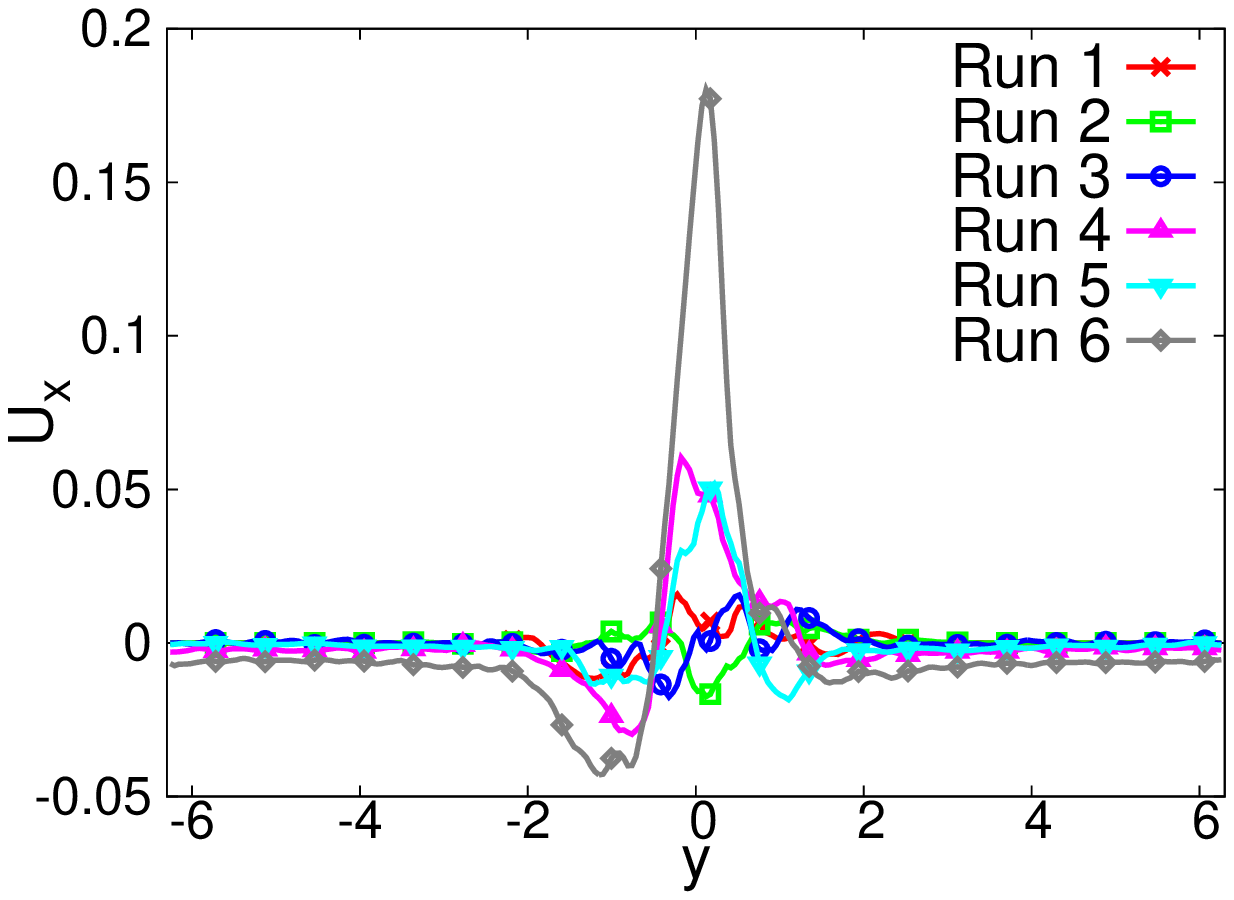}
   \caption{Mean axial velocity of each run.}
   \label{fig:3}
  \end{minipage}
 \end{tabular}
\end{figure}

Figure~\ref{fig:2} shows the time evolution of the mean axial velocity, $\left< \overline{u}_x \right>_S$, for Run 6. The mean flow is generated around $y=0$ as time elapses and is sustained in subsequent periods. This result is the same as that obtained by Yokoi and Brandenburg \cite{yb2016} in which the positive mean velocity directed to the rotation axis was generated around the positively helical region.
In the simulation performed by Yokoi and Brandenburg \cite{yb2016}, helicity is distributed as $H(y) \propto \sin (\pi y/y_0)$ (in the original study, the inhomogeneous direction is $z$), and thus the positive axial velocity emerges in $y>0$ and the negative axial velocity emerges in $y<0$. Conversely, in the present simulation, the positive helicity is driven only in a limited region around $y=0$. Hence, the positive axial mean velocity emerges only around $y=0$.
It should be noted again that the mean velocity cannot be directly generated from the external force since the direct contribution from the external force is excluded in the calculation. Hereafter, we take the time average over $20 \le t \le 30$ as well as the homogeneous plane average.
The mean axial velocity of each run is given in Fig.~\ref{fig:3}. Evidently, the positive axial mean velocity emerges only for the cases with both helicity injection and system rotation, namely Runs 4, 5, and 6. The difference between Run 3 and Runs 4 and 6 only corresponds to the existence of the helicity injection, and thus the external force with $\alpha = 0$ does not influence the induction of the axial mean velocity. This indicates that neither inhomogeneous helicity nor system rotation by themselves are sufficient to obtain the mean-flow generation.
It is interesting to note that the maximum values of the mean flows for Runs 4 and 5 are the same. This suggests that the product of the helicity and the angular velocity of system rotation determines the mean-flow generation. The mean flow profile is expected to be symmetric about $y=0$. The present result is slightly asymmetric due to the limitations of time or ensemble average. 

\begin{figure}[htbp]
\centering
\includegraphics[scale=0.72]{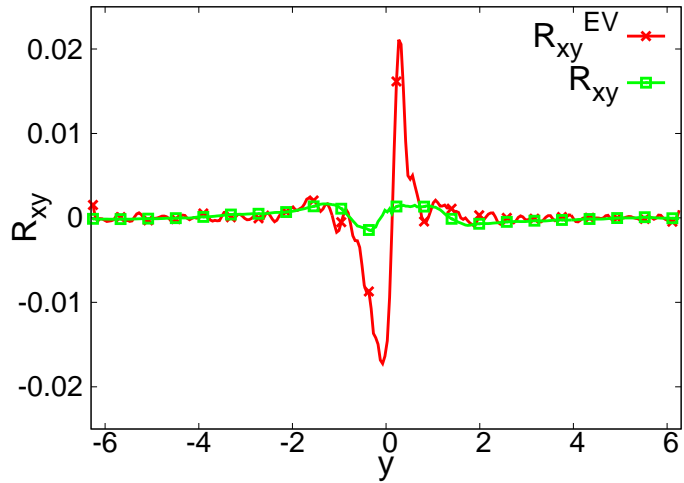}
\caption{The Reynolds stress $R_{xy}$ for Run 6. The green line with squares denotes the directly evaluated value, $R_{xy} = R_{xy}^{GS} - 2 \left< \nu_{sgs} \overline{s}_{xy} \right>$, and the red line with crosses denotes the value estimated by the eddy-viscosity model that is given by Eq.~(\ref{eq:3.9}) with $C_\nu = 0.09$.}
\label{fig:4}
\end{figure}

When the turbulent field is statistically steady, the equation for the mean axial velocity is expressed as follows:
\begin{align}
\frac{\partial U_x}{\partial t} 
= -\frac{\partial R_{xy}}{\partial y} = 0 ,
\label{eq:3.8}
\end{align}
where $R_{ij}$ satisfies $R_{ij} = R_{ij}^{GS} - 2\left<\nu_{sgs} \overline{s}_{ij} \right>$ in the framework of the eddy-viscosity representation of the SGS stress, and $R_{ij}^{GS} = \left< \overline{u}_i' \overline{u}_j' \right>$ denotes the Reynolds stress of the grid scale.
It should be noted that $\overline{u}_i'$ denotes the fluctuation of the GS velocity $\overline{u}_i$ and is defined as in Eq.~(\ref{eq:3.5}).
Equation~(\ref{eq:3.8}) gives the Reynolds stress constant in the $y$ direction. The turbulence is inactive at the upper and lower boundaries, and thus $y=\pm L_y/2$, $R_{xy}$ disappears at this point. Therefore, the solution of the mean velocity equation is $R_{xy} = 0$. The green line with squares in Fig.~\ref{fig:4} shows the profile of $R_{xy}$ for Run 6. It is nearly equal to zero although a slight non-zero value is observed around $y=0$ because the time averaging is insufficient for the statistically steady state. Here, we consider the appropriateness of the eddy-viscosity model,
\begin{align}
R_{xy} = - \nu_T \frac{\partial U_x}{\partial y} \ , \ \ 
\nu_T = C_\nu \frac{K^2}{\varepsilon},
\label{eq:3.9}
\end{align}
In Fig.~\ref{fig:4}, the profile of $R_{xy}$ estimated by Eq.~(\ref{eq:3.9}) is also plotted in the red line with crosses. It should be noted that $\nu_T$ is evaluated by using $K^{GS}$ and $\varepsilon^{SGS}$ instead of $K$ and $\varepsilon$ as $\nu_T = C\nu (K^{GS})^2/\varepsilon^{SGS}$ with $C_\nu =0.09$. This clearly indicates excessively high non-zero values around $y=0$. Since $\nu_T \neq 0$ around $y=0$, the velocity gradient must vanish in order to satisfy $R_{xy} = 0$. Therefore, the eddy-viscosity model is unable to reproduce the present result in which the mean flow is sustained around $y=0$.

In order to rectify the inadequacy of the eddy-viscosity model, let us assume the following generic expression for the model,
\begin{align}
R_{xy} = - \nu_T \frac{\partial U_x}{\partial y} + N_{xy},
\label{eq:3.10}
\end{align}
where $N_{xy}$ denotes an additional term. As shown in Fig.~\ref{fig:4}, the eddy-viscosity term, $-\nu_T \partial U_x / \partial y$, has a large positive gradient around $y=0$. In order to satisfy $R_{xy} = 0$, $N_{xy}$ must involve a large negative gradient around $y=0$ to counterbalance the eddy-viscosity term. We expect that the second term on the right-hand side of Eq.~(\ref{eq:2.14}) is a good candidate for $N_{xy}$ because the mean flow is only sustained when both the helical force and the system rotation are present.

\subsubsection{\label{sec:level3b2}Origin of the helicity effect}

\begin{figure}[htp]
\centering
\includegraphics[scale=0.72]{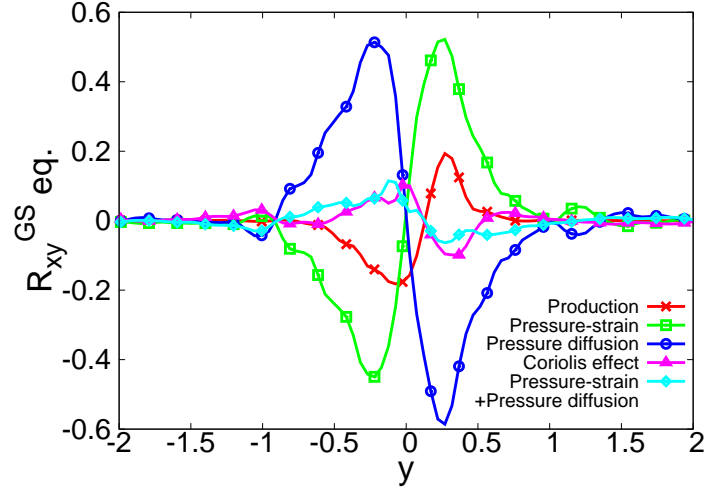}
\caption{Budget of the transport equation for $R_{xy}^{GS}$ for Run 6. The $y$ coordinate is limited to the region at $-2\le y \le 2$ where a high mean velocity exists.}
\label{fig:5}
\end{figure}

In order to investigate the origin of the additional term $N_{xy}$, we examine the transport equation for the Reynolds stress. The transport equation for $R_{xy}^{GS}$ is expressed as follows:
\begin{align}
\frac{\partial R_{xy}^{GS}}{\partial t} =
P_{xy}^{GS} + \Phi_{xy}^{GS} + \Pi_{xy}^{GS} + C_{xy}^{GS},
\label{eq:3.11}
\end{align}
where only the terms that significantly contribute to the simulation for Run 6 are included. Here $P_{xy}^{GS}$ denotes the production, $\Phi_{xy}^{GS}$ the pressure--strain correlation, $\Pi_{xy}^{GS}$ the pressure diffusion, and $C_{xy}^{GS}$ the Coriolis effect. They are respectively defined as follows:
\begin{subequations}
\begin{align}
P_{xy}^{GS} & = - \frac{2}{3} K^{GS} \frac{\partial U_x}{\partial y} 
- B_{yy}^{GS} \frac{\partial U_x}{\partial y} - B_{xz}^{GS} \frac{\partial U_z}{\partial y}, 
\label{eq:3.12a} \\
\Phi_{xy}^{GS} &
 = 2 \left< \overline{p}' \overline{s}_{xy}' \right>, 
\label{eq:3.12b} \\
\Pi_{xy}^{GS} &
 = -\frac{\partial}{\partial y} \left< \overline{p}' \overline{u}_x' \right>, 
\label{eq:3.12c} \\
C_{xy}^{GS} & = 2 R_{xz}^{GS} \Omega_x^F,
\label{eq:3.12d}
\end{align}
\end{subequations}
where $B_{ij}^{GS} = R_{ij}^{GS} - (2/3) K^{GS} \delta_{ij}$. The budget of the transport equation for $R_{xy}^{GS}$ for Run 6 is shown in Fig.~\ref{fig:5}. It should be noted that the balance of the above four terms are mostly the same for Runs 4 and 5 especially in the sense that the pressure--strain correlation and the pressure diffusion are predominant (figures are not shown here). The production term plotted in the red line with crosses exhibits a positive gradient around $y=0$. It should be noted that with respect to the production term $P_{xy}^{GS}$, the first term on the right-hand side of Eq.~(\ref{eq:3.12a}) is dominant [detailed contribution from each term in Eq.~(\ref{eq:3.12a}) is not shown here]. Thus, as shown in Sec.~\ref{sec:level2a}, the production term corresponds to the eddy-viscosity term, and it also exhibits a positive gradient around $y=0$ in Fig.~\ref{fig:4}. Based on the discussion in Sec.~\ref{sec:level3b1}, any candidate of the term corresponding to $N_{xy}$ that accounts for the sustainment of the mean velocity should exhibit a negative gradient around $y=0$. In Fig.~\ref{fig:5}, two candidates are observed, namely the pressure diffusion $\Pi_{xy}^{GS}$ (the blue line with circles) and the Coriolis effect $C_{xy}^{GS}$ (the magenta line with triangles). If the Coriolis effect corresponds to the origin of $N_{xy}$, the mean flow would be sustained for Run 3 in which  system rotation exists as well as for Runs 4, 5, and 6. Therefore, we focus on the pressure diffusion term. However, this does not deny the importance of Coriolis force in the flow generation phenomenon. As shown in Fig. 5, the Coriolis effect also contributes to the Reynolds stress in the sense that it sustains the mean flow. Additionally, the effect of the Coriolis force appears not only in the Coriolis effect but also in the pressure through the Poisson equation as discussed in the following paragraph.

\begin{figure}[htp]
\centering
\includegraphics[scale=0.72]{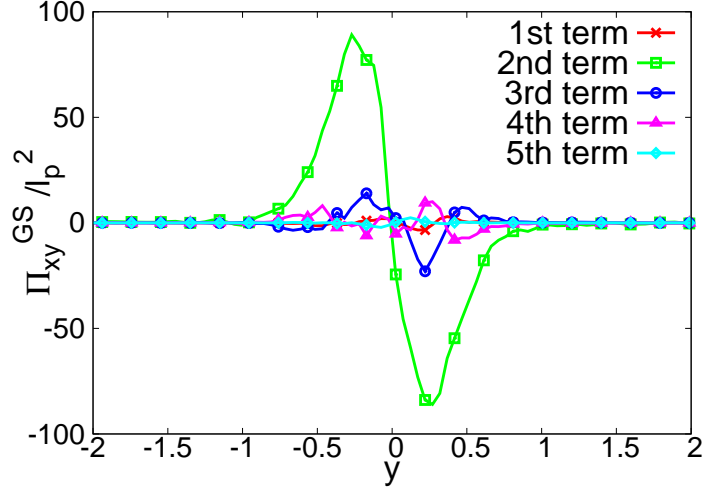}
\caption{Approximate evaluation of $\Pi_{xy}^{GS}$ for Run 6.}
\label{fig:6}
\end{figure}

In order to investigate the pressure diffusion [Eq.(\ref{eq:3.12c})], we consider the Poisson equation for the pressure fluctuation,
\begin{align}
\nabla^2 \overline{p}' & =
 - 2\overline{s}_{ab}' S_{ab} + \overline{\omega}_a' \Omega_a^*
- \overline{s}_{ab}' \overline{s}_{ab}'
+ \frac{1}{2} \overline{\omega}_a' \overline{\omega}_a' 
+ \frac{\partial^2}{\partial x_a \partial x_b} \left[ 2\left( \nu_{sgs} \overline{s}_{ab} - \left< \nu_{sgs} \overline{s}_{ab} \right> \right) \right].
\label{eq:3.13}
\end{align}
We approximate the left-hand side as
\begin{align}
\nabla^2 \overline{p}' = - \frac{\overline{p}'}{\ell_p^2},
\label{eq:3.14}
\end{align}
where $\ell_p$ denotes the length scale associated with the pressure fluctuation. Thus, the pressure diffusion $\Pi_{xy}^{GS}$ is estimated as follows:
\begin{align}
\Pi_{xy}^{GS}/\ell_p^2 & =
\frac{\partial}{\partial y} \left[
 - 2 \left< \overline{u}_x' \overline{s}_{ab}'\right> S_{ab}
 + \left< \overline{u}_x' \overline{\omega}_a' \right> \Omega_a^* 
- \left< \overline{u}_x' \overline{s}_{ab}' \overline{s}_{ab}' \right> 
+ \frac{1}{2} \left< \overline{u}_x' \overline{\omega}_a' \overline{\omega}_a' \right> 
+ \frac{\partial^2}{\partial y^2} \left( 2 \left< \overline{u}_x' \nu_{sgs} \overline{s}_{ab} \right> \right) \right] ,
\label{eq:3.15}
\end{align}
where $\ell_p$ is approximated as a constant in space for simplicity. Figure~\ref{fig:6} shows the pressure diffusion $\Pi_{xy}^{GS}$ evaluated from Eq.~(\ref{eq:3.15}) for Run 6. As shown in the figure, the second term related to the mean absolute vorticity is dominant. Thus, $\Pi_{xy}^{GS}$ is approximated as follows:
\begin{align}
\Pi_{xy}^{GS} / \ell_p^2
= \frac{\partial}{\partial y} \left( 2 \left< \overline{u}_x' \overline{\omega}_x' \right> \Omega_x^F \right)
= \frac{\partial}{\partial y} \left( \frac{2}{3} H^{GS} \Omega_x^F \right),
\label{eq:3.16}
\end{align}
and this includes $|\Omega_i| \ll |2\Omega_i^F|$ and $\left<\overline{u}_x' \overline{\omega}_x' \right> = \left<\overline{u}_y' \overline{\omega}_y' \right> = \left<\overline{u}_z' \overline{\omega}_z' \right> = H^{GS} /3$. This indicates that the helicity gradient and the system rotation may account for the pressure diffusion that contributes to the mean velocity sustainment. A model expression of the pressure diffusion that is similar to Eq.~(\ref{eq:3.16}) is also analytically obtained with the aid of the TSDIA \cite{tsdia}. A brief introduction of the theory and the detailed calculation are given in Appendix~\ref{sec:a}. The result is
\begin{align}
\Pi_{ij} & = \frac{1}{3} \left[ \frac{\partial}{\partial x_j} \left( L^2 H 2\Omega_i^F \right)
+ \frac{\partial}{\partial x_i} \left( L^2 H 2\Omega_j^F \right) \right] 
+ \text{(non-helical term)} + O(|u^{(00)}|^3), 
\label{eq:3.17}
\end{align}
where $L$ denotes the length scale related to the energy containing eddy and $u^{(00)}$ is the lowest-order velocity corresponding to homogeneous isotropic turbulence defined in Eq.~(\ref{eq:a4}). This model expression for the pressure diffusion is in good agreement with  Eq.~(\ref{eq:3.16}).

In Fig.~\ref{fig:5}, the pressure--strain correlation $\Phi_{ij}^{GS}$ (the green line with squares) also significantly contributes to the Reynolds-stress transport. One might consider that the pressure diffusion and the pressure--strain correlation cancel each other. However, the sum of these two terms (as denoted by the cyan line with diamonds in Fig.~\ref{fig:5}) contributes to exhibit a negative gradient around $y=0$ and plays the same role as the pressure diffusion itself. This tendency is also theoretically demonstrated as follows. The model expression of the pressure--strain correlation $\Phi_{ij}$ is obtained with the aid of the TSDIA \cite{tsdia}, and it is possible to analytically examine the balance of the two terms. The analytical result of the pressure--strain correlation is as follows:
\begin{align}
\Phi_{ij} & = - \frac{3}{10} \left[ \frac{\partial}{\partial x_j} \left( L^2 H 2 \Omega_i^F \right)
   + \frac{\partial}{\partial x_i} \left( L^2 H 2\Omega_j^F \right) \right]_D 
+ \text{(non-helical term)} + O(|u^{(00)}|^3).
\label{eq:3.18}
\end{align}
Although the helicity effect of the pressure--strain correlation $\Phi_{ij}$ has the sign opposite to that of the pressure diffusion $\Pi_{ij}$, its magnitude is slightly smaller. Thus the sum of $\Pi_{ij}$ and $\Phi_{ij}$,
\begin{align}
\Pi_{ij} + \Phi_{ij} & = 
\frac{1}{30} \left[ \frac{\partial}{\partial x_j} \left( L^2 H 2\Omega_i^F \right)
+ \frac{\partial}{\partial x_i} \left( L^2 H 2\Omega_j^F \right) \right] 
+ \text{(non-helical term)} + O(|u^{(00)}|^3),
\label{eq:3.19}
\end{align}
contributes in the same manner as the pressure diffusion $\Pi_{ij}$ and sustains the mean flow.

\section{\label{sec:level4}Correspondence of the pressure diffusion to the helicity model}

The results indicated that the pressure diffusion plays an important role in the sustainment of the mean velocity  in inhomogeneous helical turbulence. This fact contradicts the assumption for the derivation of the model for the Reynolds stress as given in Sec.~\ref{sec:level2a}. In the current construction, the flow is assumed to be quasi-homogeneous for the diffusion to be neglected. However, the effect of the pressure diffusion is required to improve the Reynolds-stress model for inhomogeneous helical turbulence. As shown in Sec.~\ref{sec:level3b2}, the effect of helicity is explicitly incorporated in the pressure diffusion term for the Reynolds-stress transport equation. Here, we add the helicity effect that originates from the pressure diffusion term to the LRR model \cite{lrr1975} as follows:
\begin{align}
\begin{split}
& \Phi_{ij} + \left[ \Pi_{ij} \right]_D
= \Phi_{ij}^{LRR} + C_{PH} \Gamma_{ij},
\label{eq:4.1}
\end{split}
\end{align}
where $\Phi_{ij}^{LRR}$ denotes the LRR model given by Eq.~(\ref{eq:2.10}), $C_{PH}$ is a positive constant, and
\begin{align}
\Gamma_{ij} = 
 \left[ \frac{\partial}{\partial x_j} \left( \frac{K^3}{\varepsilon^2} H \Omega_i^* \right)
+ \frac{\partial}{\partial x_i} \left( \frac{K^3}{\varepsilon^2} H \Omega_j^* \right) \right]_D .
\label{eq:4.2}
\end{align}
Here, the length scale that corresponds to $\ell_p$ in Eq.~(\ref{eq:3.14}) or $L$ in Eq.~(\ref{eq:3.17}) is expressed in terms of $K$ and $\varepsilon$. Thus, the Reynolds-stress equation is re-expressed as follows:
\begin{align}
\frac{\mathfrak{D} B_{ij}}{\mathfrak{D} t} & =
- C_{S1} \frac{\varepsilon}{K} B_{ij}
- \left( \frac{4}{3} - C_{R1} \right) K S_{ij} 
+ C_{PH} \Gamma_{ij} \nonumber \\
& \hspace{1.2em}
- \left( 1 - C_{R2} \right) \left[ B_{i\ell} S_{\ell j} + B_{j\ell} S_{\ell i} \right]_D 
- \left( 1 - C_{R3} \right) \left( B_{i\ell} \Omega_{\ell j}^* + B_{j\ell} \Omega_{\ell i}^* \right),
\label{eq:4.3}
\end{align}
The third term on the right-hand side denotes the only difference between Eqs.~(\ref{eq:2.11}) and (\ref{eq:4.3}). Thus, the model expression corresponding to Eq.~(\ref{eq:2.13}) is given as follows:
\begin{align}
B_{ij} & = - 2 C_\nu \frac{K^2}{\varepsilon} S_{ij}
+ C_\gamma \frac{K}{\varepsilon} \Gamma_{ij}
+ \cdots,
\end{align}
where $C_\gamma = C_{PH} / C_{S1}$. The second term is significantly similar to Eq.~(\ref{eq:2.14}) obtained by Yokoi and Yoshizawa \cite{yy1993}. Hence, the helicity model given by Eq.~(\ref{eq:2.14}) can trace part of its origin to the pressure diffusion in inhomogeneous helical turbulence in a rotating system.

\section{\label{sec:level5}Conclusions}

The mechanism of the mean-flow generation and its relationship to the turbulent helicity were investigated by using the numerical simulation of a rotating inhomogeneous turbulence. In the simulation, an external forcing was applied to inject turbulent energy and helicity and the rotation axis was perpendicular to the inhomogeneous direction. The initial mean velocity and the mean part of the external force were set to zero, and this implies that it is not possible to directly excite the mean flow by the external forcing. The results showed that the mean flow is generated and sustained only when both helical forcing and system rotation exist. The flow-generation phenomenon originates from both the turbulent helicity and the rotational motion of fluid. 

The usual eddy-viscosity model is unable to reproduce the mean-flow generation observed in the simulation, and therefore an additional term is needed to explain the phenomenon. In order to explore candidates for the additional term, the budget of the Reynolds-stress transport equation was investigated. The results suggested that the pressure diffusion significantly influences the sustainment of the mean flow. The approximation to the Poisson equation for the pressure fluctuation was used to obtain an expression for the pressure diffusion in terms of the turbulent helicity and the angular velocity of the system rotation. The effect of helicity in relation to the pressure diffusion term was considered to obtain a model for the Reynolds stress, and the obtained model is considerably similar to the one obtained by Yokoi and Yoshizawa \cite{yy1993}. The model implies that the inhomogeneity of helicity plays a crucial role in rotating turbulence such as the momentum transport due to turbulence in the low-latitude region of a rotating sphere \cite{duarteetal2016,ssd2014}.

\appendix

 \makeatletter
 \renewcommand{\theequation}{
 \thesection\arabic{equation}}
 \@addtoreset{equation}{section}
 \makeatother

\section{\label{sec:a}{Analytical modeling of helicity effect on the pressure-related terms}}

The effect of helicity on the pressure diffusion and the pressure--strain correlation is estimated by using the TSDIA \cite{tsdia} that corresponds to a closure scheme for inhomogeneous turbulence. In this formalism, the fast variables $(\bm{\xi}, \tau)$ and slow variables $(\bm{X},T)$ are introduced for space and time variables with a scale parameter $\delta$,
\begin{align}
\bm{\xi} = \bm{x}, \ \ \tau = t, \ \ 
\bm{X} = \delta \bm{x}, \ \ T = \delta t.
\label{eq:a1}
\end{align}
We assume that the fluctuation fields depend on both the fast and slow variables while mean fields depend on only the slow variables, and this is expressed as follows:
\begin{align}
q = Q (\bm{X};T) + q' (\bm{\xi},\bm{X};\tau,T),
\label{eq:a2}
\end{align}
where $q = (u_i,p)$. The space and time derivatives are then re-expressed as follows:
\begin{align}
\frac{\partial}{\partial x_i} = \frac{\partial}{\partial \xi_i} + \delta \frac{\partial}{\partial X_i}, \ \ 
\frac{\partial}{\partial t} = \frac{\partial}{\partial \tau} + \delta \frac{\partial}{\partial T} .
\label{eq:a3}
\end{align}
We expand $q'$ in powers of $\delta$ and the rotation parameter $\Omega^F$ as follows \cite{yy1993},
\begin{align}
q' (\bm{\xi},\bm{X}; \tau, T) = \sum_{n,m=0}^\infty \delta^n |\Omega^F|^m q^{(nm)} (\bm{\xi},\bm{X}; \tau, T).
\label{eq:a4}
\end{align}
The $O(\delta^0 |\Omega^F|^0)$ field corresponds to the homogeneous turbulence.
In this formalism, the effects of inhomogeneity and anisotropy and the effects of rotation are systematically incorporated in the higher-order fields, $O(\delta^n)$ and/or $O(|\Omega^F|^m)$ with $n \ge 1$ or $m \ge 1$, in a perturbational manner. Subsequently, the Fourier transformation is applied to the fast variables. With respect to the lowest-order field, we assume the following statistical properties,
\begin{align}
& \left< \tilde{u}_i^{(00)} (\bm{k},\bm{X}; \tau, T) \tilde{u}_j^{(00)} (\bm{k}',\bm{X}; \tau' , T) \right> \nonumber \\
& = \left[ D_{ij} (\bm{k}) \frac{E^0 (k,\bm{X}; \tau, \tau', T)}{4 \pi k^2} 
+ \frac{i}{2} \frac{k_\ell}{k^2} \epsilon_{ij\ell} \frac{E_H^0 (k,\bm{X}; \tau, \tau', T)}{4 \pi k^2} \right] \delta (\bm{k}+\bm{k}') , 
\label{eq:a5}
\end{align}
where $D_{ij} (\bm{k}) = \delta_{ij} - k_i k_j / k^2$ and $E^0$ and $E_H^0$ denote the spectral functions of the turbulent energy and helicity, respectively, which satisfy the following expression:
\begin{align}
\frac{1}{2} \left< u_i^{(00)} u_i^{(00)} \right> & = \int_0^\infty \mathrm{d} k E^0 (k, \bm{X}; \tau, \tau, T) , 
\label{eq:a6} \\
\left< u_i^{(00)} \omega_i^{(00)} \right> & = \int_0^\infty \mathrm{d} k E_H^0 (k, \bm{X}; \tau, \tau, T) .
\label{eq:a7}
\end{align}
The higher-order fields are solved by using the Green's function $G_{ij}$ of the lowest-order velocity equation that corresponds to a homogeneous turbulent field. The statistical average of the Green's function is given as follows:
\begin{align}
\left< G_{ij} (\bm{k}, \bm{X}, \tau, \tau', T) \right> = D_{ij} (\bm{k}) G (k, \bm{X}; \tau, \tau', T).
\label{eq:a8}
\end{align}
Following the calculation we replace $E^0$ and $E_H^0$ by $E$ and $E_H$, respectively, namely we renormalize the lowest-order velocity correlations by the exact correlations. Up to $O(\delta |\Omega^F|)$, the Reynolds stress, the pressure diffusion, and the pressure--strain correlation are calculated as follows:
\begin{align}
B_{ij} & = \left[ \left< u_i^{(00)} u_j^{(00)} \right> + \left< u_i^{(01)} u_j^{(00)} \right>
+ \left< u_i^{(10)} u_j^{(00)} \right> + \left< u_i^{(11)} u_j^{(00)} \right> + ( i \leftrightarrow j ) \right]_D \nonumber \\
& = - 2 \nu_T S_{ij} + \left[ \chi_i 2\Omega^F_j + \chi_j 2 \Omega^F_i \right]_D,
\label{eq:a9} \\
\Pi_{ij} & = \delta \frac{\partial}{\partial X_j} \left[
   \left< u_i^{(00)} p^{(00)} \right> + \left< u_i^{(01)} p^{(00)} \right> 
+ \left< u_i^{(00)} p^{(01)} \right> \right] + ( i \leftrightarrow j ) \nonumber \\
& = \frac{1}{3} \left[ \frac{\partial}{\partial x_j} \left( L^2 H 2\Omega_i^F \right)
+ \frac{\partial}{\partial x_i} \left( L^2 H 2\Omega_j^F \right) \right] 
+ \text{(non-helical term)} + O(|u^{(00)}|^3),
\label{eq:a10} \\
\Phi_{ij} & = 2 \left< s_{ij}^{(00)} p^{(00)} \right>
 + 2 \left< s_{ij}^{(01)} p^{(00)} \right> + 2 \left< s_{ij}^{(00)} p^{(01)} \right> \nonumber \\
& \hspace{1em} + \delta \left[ 2 \left< s_{ij}^{(10)} p^{(00)} \right>
+ 2 \left< s_{ij}^{(00)} p^{(10)} \right> 
+ 2 \left< s_{ij}^{(11)} p^{(00)} \right> \right. \nonumber \\
& \hspace{3em} \left. + 2 \left< s_{ij}^{(10)} p^{(01)} \right> 
+ 2 \left< s_{ij}^{(01)} p^{(10)} \right> 
+ 2 \left< s_{ij}^{(00)} p^{(11)} \right> \right] \nonumber \\
& = - \frac{3}{10} \left[ \frac{\partial}{\partial x_j} \left( L^2 H 2 \Omega_i^F \right)
   + \frac{\partial}{\partial x_i} \left( L^2 H 2\Omega_j^F \right) \right]_D 
+ \text{(non-helical term)} + O(|u^{(00)}|^3),
\label{eq:a11}
\end{align}
where
\begin{subequations}
\begin{align}
\nu_T & = \frac{7}{15} \int_0^\infty \mathrm{d} k \int_{-\infty}^\tau \mathrm{d} \tau' \ 
G (k, \tau, \tau') E (k, \tau, \tau'), 
\label{eq:a12a} \\
\chi_i & = \eta \frac{\partial H}{\partial x_i} =
 \frac{1}{30} \int_0^\infty \mathrm{d} k \ k^{-2} \int_{-\infty}^\tau \mathrm{d} \tau' \ 
G (k, \tau, \tau') \frac{\partial E_H (k,\tau,\tau')}{\partial x_i} ,
\label{eq:a12b} \\
L^2 H & = \int_0^\infty \mathrm{d}k \ k^{-2} E_H (k, \tau, \tau), 
\label{eq:a12c} \\
s_{ij}^{(0n)} & = \frac{1}{2} \left( \frac{\partial u_i^{(0n)}}{\partial \xi_j} + \frac{\partial u_j^{(0n)}}{\partial \xi_i} \right), 
\label{eq:a12d} \\ 
s_{ij}^{(1n)} & = \frac{1}{2} \left( \frac{\partial u_i^{(1n)}}{\partial \xi_j} + \frac{\partial u_j^{(1n)}}{\partial \xi_i} \right) 
+ \frac{1}{2} \left( \frac{\partial u_i^{(0n)}}{\partial X_j} + \frac{\partial u_j^{(0n)}}{\partial X_i} \right) .
\label{eq:a12e}
\end{align}
\end{subequations}

\section{\label{sec:b}{Details of the external force}}

The vector potential $\overline{\psi}_i$ in Eq.~(\ref{eq:3.4}) is obtained by solving the following time evolution equation:
\begin{align}
\overline{\psi}_i (t+\Delta t) =
\left( 1 - \frac{\Delta t}{\tau} \right) \overline{\psi}_i (t) + \frac{\Delta t}{\tau} r_i ,
\label{eq:b1}
\end{align}
where $\tau = 50\Delta t$ and the vector $r_i$ is generated by using a random variable. This corresponds to the Ornstein--Uhlenbeck process with the variance of $\sigma_{OU} = \sqrt{\Delta t / (2\tau)}$ when $r_i$ denotes the Gaussian random variable \cite{vfj2010}. If the weighting function $g(y)$ in Eq.~(\ref{eq:3.4}) is constant in space, then the one-point two-time correlation of the external force is expressed as \cite{ouforcing}
\begin{align}
\left< \overline{f}_i (\bm{x}, t) \overline{f}_j (\bm{x}, s) \right> \propto
\delta_{ij} \mathrm{e}^{-(t-s)/\tau} .
\label{eq:b2}
\end{align}
The amplitude of the random vector $r_i$ is determined by considering the power and helicity spectra of $\overline{f}_i$, $E^{ex} (k)$ and $E_H^{ex}(k)$, as follows:
\begin{subequations}
\begin{align}
\frac{1}{2} \left< r_i r_i \right> & = \int_0^\infty \mathrm{d} k \ k^{-2} E^{ex} (k) , 
\label{eq:b3a} \\
\left< r_i \epsilon_{ij\ell} \frac{\partial r_\ell}{\partial x_j} \right> & = 
\int_0^\infty \mathrm{d} k \ k^{-2} E_H^{ex} (k) .
\label{eq:b3b}
\end{align}
\end{subequations}

\bibliography{ref}

\end{document}